\documentstyle[12pt]{article}

\begin{document}

\title{Dark Matter: a Challenge to Standard Gravity or a
Warning?\footnote{UCONN 97-01, January 1997, to appear in proceedings of
18th Texas
Symposium on Relativistic Astrophysics, Chicago, December 1996.}}

\author{\normalsize{Philip D. Mannheim} \\
\normalsize{Department of Physics,
University of Connecticut, Storrs, CT 06269} \\
\normalsize{mannheim@uconnvm.uconn.edu} \\}

\date{}

\maketitle

\begin{abstract}
We suggest that the conventional need for overwhelming amounts of astrophysical
dark matter should be regarded as a warning to standard gravity rather than
as merely a challenge to it, and show that the systematics of galactic rotation
curve data can just as readily point in the direction of the equally covariant
conformal gravity alternative. In particular we identify an apparent imprint of
the Hubble flow on those data, something which while quite natural to conformal
gravity is not at all anticipated in the standard gravitational paradigm.
\end{abstract}

\noindent
While the main thrust of current research is to meet the challenge of
determining the precise nature and form of the dark matter which is widely
thought to dominate the universe on large distance scales, nonetheless, given
so
startling a requirement as this, it is not inappropriate to ask whether this
very need for dark matter, and in such copious proportions, might not instead
be
a warning that the standard Newton-Einstein theory might not in fact be the
right one for gravity. Now while the standard theory was first established on
solar system distance scales, for the moment its extrapolation to much larger
distances is precisely just that, since there is not yet a single independent
verification of Newton's Law of Gravity on galactic or larger
distance scales which does not involve an appeal to dark matter, to thus show
the complete circularity of the very reasoning which leads to dark matter in
the
first place. Thus at the present time observation can only mandate that gravity
be a covariant theory whose metric reproduces the familiar Schwarzschild
systematics on solar system distance scales, with (as noted long ago by
Eddington) this actually being readily achievable in theories of gravity of
order higher than the standard second order one. Since such higher order
theories would however then also depart from Schwarzschild at larger distances
\cite{MannheimandKazanas1994} their phenomenological candidacy is immediate. In
fact, motivated by the underlying conformal invariance of the three other
fundamental interactions, Mannheim and Kazanas considered the candidacy of one
explicit higher order theory, viz. fourth order conformal gravity, and
found \cite{MannheimandKazanas1989} that outside of a static, spherically
symmetric system such as a star the exact metric takes the form
$-g_{00}= 1/g_{rr}=1-2\beta^{*}/r + \gamma^{*} r$, to thus precisely recover
the
Schwarzschild metric on small enough distances while both generalizing it and
departing from it on much larger ones.

\noindent
While higher order theories of gravity such as conformal gravity thus yield
potentials which then dominate over the Newtonian one at large distances just
as
desired, the very fact that they do so entails that one is now no longer able
to
ignore the potentials due to distant matter sources outside of individual
gravitational systems such as galaxies. Thus once we depart from the second
order theory, we immediately transit into a world where we have to consider
effects due to matter not only inside but also outside of individual systems,
and we are thus led \cite{Mannheim1995a,Mannheim1996a,Mannheim1996b} to look
for
both local and global imprints on galactic rotation curve data, this being a
quite radical (and quite Machian) conceptual departure from the standard purely
local Newtonian world view.

\noindent
To isolate such possible global imprints, it is instructive
\cite{Mannheim1995a,Mannheim1996a,Mannheim1996b} to look at the
centripetal accelerations of the data points farthest from galactic centers.
In particular, for a large set of galaxies whose rotation curve data are
regarded as being particularly characteristic of the pattern of deviation from
the luminous Newtonian expectation that has so far been obtained, it was found
that these farthest centripetal accelerations could all be parameterized by
the universal three component relation
$(v^2/R)_{last}=\gamma_0c^2/2+\gamma^{*}N^{*}c^2/2 +\beta^{*}N^{*}c^2/R^2$
where $\gamma_0=3.06\times 10^{-30}$ cm$^{-1}$,
$\gamma^{*}=5.42\times 10^{-41}$ cm$^{-1}$, $\beta^{*}=1.48\times 10^5$ cm, and
where $N^{*}$ is the total amount of visible matter (in solar mass units) in
each galaxy. Since the luminous Newtonian contribution is decidedly non-leading
at the outskirts of galaxies, we thus uncover the existence of two linear
potential terms which together account for the entire measured departure
from the luminous Newtonian expectation (not only for these farthest points but
even \cite{Mannheim1996a,Mannheim1996b} for all the other (closer in)
data points as well in fact). Now while the $\gamma^{*}N^{*}c^2/2$ term can
immediately be identified as the net galactic contribution due to the linear
$\gamma^{*}c^2R/2$ potentials of all the $N^{*}$ stars in each galaxy, the
inferred $\gamma_0c^2/2$ term is on a quite different footing since
it is independent of the mass content $N^{*}$ of each of the individual
galaxies. Moreover, since numerically $\gamma_0$ is found to have a magnitude
of
order the inverse Hubble radius, we can thus anticipate that
it must represent a universal global effect generated by the matter outside of
each galaxy (viz. the rest of the matter in the universe), and thus not be
associated with any local dynamics within individual galaxies at all.

\noindent
The emergence of the $\gamma_0c^2/2$ term, a term which may be thought of as
being a universal acceleration, immediately raises some questions. First, if it
is an acceleration, then with respect to which frame is the acceleration - and
no matter which particular one (the Hubble flow itself being the only apparent
covariant possibility), how could it possibly be universal for each and every
galaxy. And moreover, how could $\gamma_0$ be related to the Hubble radius at
all since the Hubble parameter is not a static, time independent quantity. As
we
shall see, conformal gravity provides
\cite{Mannheim1995a,Mannheim1996a,Mannheim1996b}
answers to all these questions. Specifically, it was noted quite early on
\cite{MannheimandKazanas1989} that the general coordinate transformation
$r=\rho/(1-\gamma_0 \rho/4)^2$, $t = \int d\tau / R(\tau)$ transforms the
metric
$ds^2=(1+\gamma_0 r)c^2dt^2-dr^2 /(1+\gamma_0 r)-r^2d\Omega$ into
$ds^2=\Lambda(\rho, \tau)\{c^2 d\tau^2 - R^2(\tau) (d\rho^2 + \rho^2 d\Omega)
/(1-\rho^2\gamma_0^2/16)^2\}$ (where the conformal factor $\Lambda(\rho, \tau)$
is given by $(1+\rho\gamma_0/4)^2 / R^2(\tau)(1-\rho\gamma_0/4)^2$), to yield a
metric which is conformal to a Robertson-Walker metric with scale factor
$R(\tau)$ and explicitly negative 3-space scalar curvature $k=-\gamma_0^2/4$,
with this metric in fact being none other than that explicitly
found \cite{Mannheim1992,Mannheim1995b} in conformal cosmology where only an
open universe with explicitly negative $k$ is realizable. Now, in a geometry
which is both homogeneous and isotropic about all points, any observer located
at the center of any comoving galaxy can serve as the origin for the coordinate
$\rho$. Thus in his own local rest frame each such comoving observer is able to
make the above coordinate transformation with the use of his own particular
$\rho$, to then find that in his own frame the entire Hubble flow then acts as
a
universal linear potential coming directly from the spatial curvature of the
universe; with galactic test particles then experiencing a universal
acceleration $\gamma_0c^2/2$ which is explicitly generated by
the rest of the matter in the universe. Thus we see that rather then being
an acceleration with respect to the Hubble flow (a non-relativistic notion),
explicitly because of relativity, the universal acceleration in fact emerges as
an intrinsic property of the Hubble flow itself as seen in each comoving
observer's rest frame. Moreover, with it also emerging as the spatial
curvature,
it is then also a nicely time independent quantity, with the numerical
determination of $\gamma_0$ given above actually yielding an explicit value for
both the sign and magnitude of the spatial curvature of the universe, something
which years of intensive work has yet to accomplish in the standard theory.
Thus
to conclude, it would appear from our analysis that there is something
explicitly global at play in galactic dynamics, something not only quite
suggestive of conformal gravity but also seemingly somewhat foreign to the
standard gravitational paradigm. This work has been supported in part by the
Department of Energy under grant No. DE-FG02-92ER40716.00.

\end{document}